\documentclass{webofc}
\usepackage[varg]{txfonts}   
\usepackage{color}
\usepackage{float}
\begin{document}
\title{Charming loops in exclusive rare FCNC $B$-decays}
\author{
\firstname{Dmitri} \lastname{Melikhov}\inst{1,2,3}\fnsep\thanks{\email{dmitri_melikhov@gmx.de}}
}
\institute{D.~V.~Skobeltsyn Institute of Nuclear Physics, M.~V.~Lomonosov Moscow State University, 119991, Moscow, Russia
\and
           Institute for High Energy Physics, Austrian Academy of Sciences, Nikolsdorfergasse 18, A-1050 Vienna, Austria
\and
           Faculty of Physics, University of Vienna, Boltzmanngasse 5, A-1090 Vienna, Austria
          }

%
%
%

\abstract{Rare $B$-decays induced by flavour-changing neutral currents (FCNC) is one of the promising candidates for probing 
physics beyond the Standard model. However, for identifying potential new physics from the data, reliable control 
over QCD contributions is necessary. We focus on one of such QCD contributions -- the charming loops -- that potentially can lead to 
difficulties in disentangling new physics effects from the observables and discuss the possibility to gain control over theoretical 
predictions for charming loops.}
\maketitle

\section{Introduction}
\label{Sec_introduction}
The interest in the contributions of charm to rare FCNC decays of the $B$-mesons is to a great extent motivated by the fact 
that virtual charm-quark loops, including charmonia states which appear in the physical region of several FCNC $B$-decay,  
have a strong impact on the $B$-decay observables \cite{neubert} thus providing a ``noise'' for the extraction of possible 
new physics effects.
\begin{figure}[h]
\centerline{\includegraphics[width=12cm]{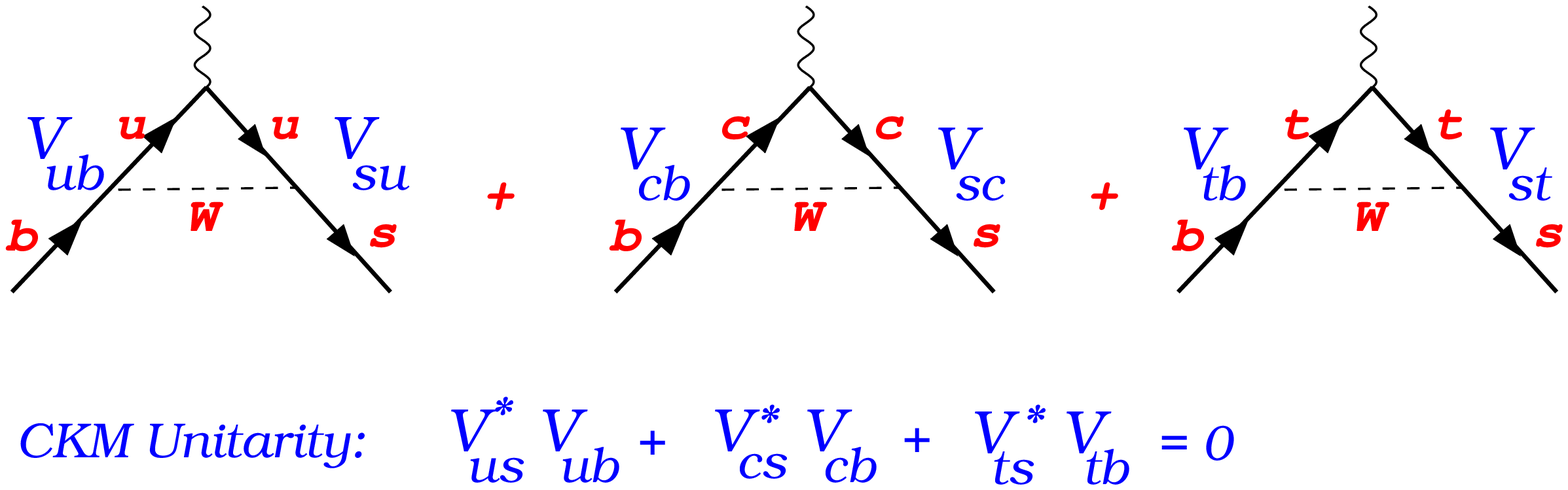}}
\caption{\label{Fig:1a}
The amplitude of a rare FCNC $b\to s \gamma$ transition in the SM. The process proceeds via the contribution of 
positive-charged quarks in the loops. Recall that the unitarity of the CKM matrix leads to the cancellation of 
the leading ultraviolet divergences in the sum of the penguin diagrams.} 
\end{figure}
FCNC decays are forbidden in the SM and proceed through loop contributions, see Figure \ref{Fig:1a}. 
The $u$-quark contribution to the CKM unitarity is strongly suppressed, leading to  
\begin{eqnarray}
V_{tb}V^*_{ts}\simeq - V_{cb}V^*_{cs},
\end{eqnarray}
so the contributions of the top and the charm have approximately the same CKM strength, while the $u$-quark contribution in the loop 
may be neglected.

\subsection{Top-quark contribution to FCNF $B$-decays}
\begin{figure}[b]
\centerline{\includegraphics[width=12cm]{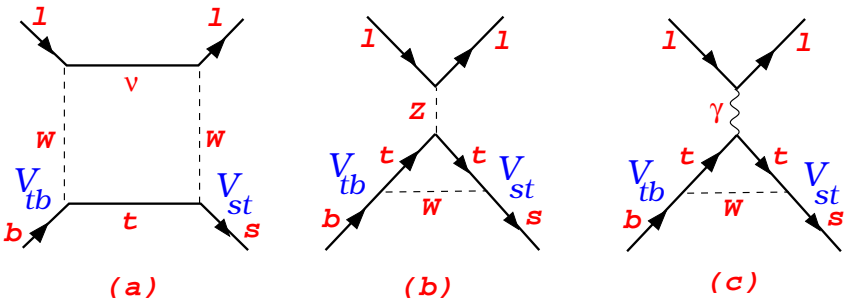}}
\caption{\label{Fig:1c}
The amplitude of the rare FCNC $b\to s l^+l^-$ transition in the SM. After integrating our $t$, $W$, and $Z$, diagrams (a) and (b) are 
reduced to local operators $O_{9V}$ and $O_{10A}$, whereas diagram (c) leads to the term proportional to the operator $O_{7\gamma}$.} 
\end{figure}

In $B$-decays, the characteristic energy scale $\mu\sim m_b$ is much lower than the masses of the $t$-quark and the $W$ and $Z$ 
bosons, therefore these heavy degrees of freedom may be integrated out \cite{il,Ben,Burasa,Burasb}. 
\begin{figure}[h]
\centerline{\includegraphics[width=8cm]{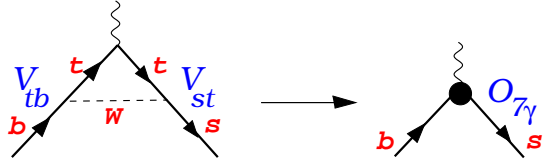}}
\caption{\label{Fig:1b}
The $t$-quark electromagnetic penguin in the amplitude of a rare FCNC $b\to s \gamma$ transition in the SM. 
Integrating out the $t$-quark and $W$-boson contributions lead to the local operator 
$O_{7\gamma}=\bar s\,\sigma_{\mu\nu}\left (1+\gamma_5\right )b \cdot F^{\mu\nu}$.} 
\end{figure}
As the result, e.g. for the top contribution to the effective Hamiltonian describing the $b\to s\gamma$ decay, Fig.~\ref{Fig:1b}, 
one finds the following expression:
\begin{eqnarray}
\label{top}
H_{\rm eff}^{b\to s\gamma}({\rm top})\, = - \frac{G_{F}}{\sqrt2}\, V_{tb}V^*_{ts}\, 
C_{7\gamma}(\mu)\,\frac{e}{8\pi^2}\, m_b \cdot
\bar s\,\sigma_{\mu\nu}\left (1+\gamma_5\right )b \cdot F^{\mu\nu}.
\end{eqnarray}
To calculate the effective Hamiltonian for the $b\to sl^+l^-$ transition, one takes into account the contribution generated by the 
electromagnetic penguin operator (\ref{top}), and adds the contribution of box and penguin diagrams described by operators $O_{9V}$ and $O_{10A}$
(Fig.~\ref{Fig:1c}) yielding 
\begin{eqnarray}
\label{topll}
&&H_{\rm eff}^{b\to s l^{+}l^{-}}({\rm top})\,=\, 
{\frac{G_{F}}{\sqrt2}}\, {\frac{\alpha_{\rm em}}{2\pi}}\, 
V_{tb}V^*_{ts}\, 
\left[\,-2im_b\, {\frac{C_{7\gamma}(\mu)}{q^2}}\cdot
\bar s\sigma_{\mu\nu}q^{\nu}\left (1+\gamma_5\right )b
\cdot{\bar l}\gamma^{\mu}l \right.
\nonumber\\
&&\qquad\qquad\left.+\, 
C_{9V}(\mu)\cdot\bar s \gamma_{\mu}\left (1\, -\,\gamma_5 \right)   b 
\cdot{\bar l}\gamma^{\mu}l \, +\, 
C_{10A}(\mu)\cdot\bar s   \gamma_{\mu}\left (1\, -\,\gamma_5 \right) b 
\cdot{\bar l}\gamma^{\mu}\gamma_{5}l \right]. 
\end{eqnarray} 

\subsection{Charm-quark contribution to FCNC $B$-decays}
The $c$-quark is dynamical at the scale $\mu\simeq m_b$, so at most one can integrate out the $W$-boson contribution and come to the 
four-fermion effective Hamiltonian 
\begin{figure}[t]
\centerline{\includegraphics[width=9cm]{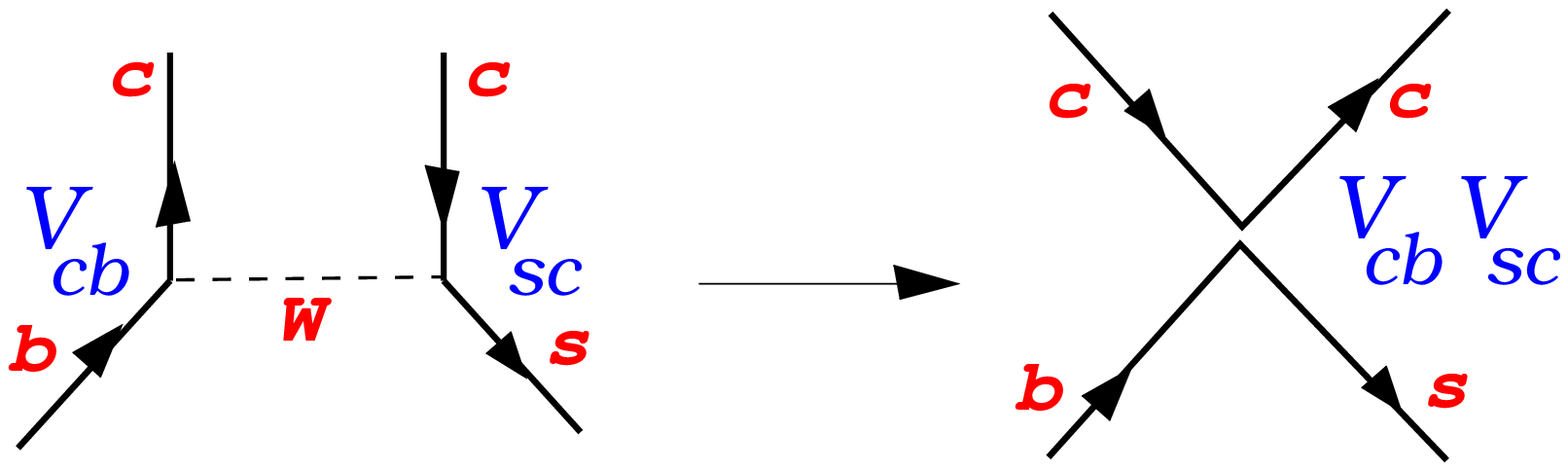}}
\caption{\label{Fig:1d}
The amplitude of the rare FCNC $b\to s l^+l^-$ transition in the SM. After integrating our $t$, $W$, and $Z$, diagrams (a) and (b) are 
reduced to local operators $O_{9V}$ and $O_{10A}$, whereas diagram (c) is proportional to $O_{7\gamma}$.} 
\end{figure}
\begin{equation}
\label{charm}
H_{\rm eff}^{b\to s\bar cc} = - \frac{G_{F}}{\sqrt2}\, V_{cb}V^*_{cs}\, 
\left\{C_{1}(\mu){\cal O}_1 + C_{2}(\mu){\cal O}_2\right\}
\end{equation}
with 
\begin{eqnarray}
{\cal O}_1=\bar s^j \gamma_\mu(1-\gamma_5)c^i \,\bar c^i \gamma^\mu(1-\gamma_5)b^j, \qquad 
{\cal O}_2=\bar s^i \gamma_\mu(1-\gamma_5)c^i \,\bar c^j \gamma^\mu(1-\gamma_5)b^j, 
\end{eqnarray}
The SM Wilson coefficients at the scale $\mu_0=5$ GeV have the values [corresponding to $C_2(M_W)=-1$] \cite{Burasa,Burasb}: 
$C_1(\mu_0)=0.241$, 
$C_2(\mu_0)=-1.1$, 
$C_7(\mu_0)=0.312$, 
$C_{9V}(\mu_0)=-4.21$, $C_{10A}(\mu_0)=4.41$.
\subsubsection{Factorizable charm contributions}
Using Eq.~(\ref{charm}), factorizable charm contributions to the $b\to s l^+l^-$ amplitude reads \cite{hidr}: 
\begin{eqnarray}
\label{charmll}
&&H_{\rm eff}^{b\to s l^{+}l^{-}}({\rm charm, fact}) =\, 
{\frac{G_{F}}{\sqrt2}}\, {\frac{\alpha_{\rm em}}{2\pi}}\, 
V_{cb}V^*_{cs}\, (C_2+3C_1)\,g_{cc}(\hat m_c^2,\hat q^2)\nonumber\\
&&\qquad\qquad\qquad\qquad\qquad\qquad\times \left(\bar s \gamma_{\mu}\left (1\, -\,\gamma_5 \right)b\right) \;\left({\bar l}\gamma^{\mu}l\right),
\end{eqnarray} 
where $g_{cc}(\hat m_c^2,\hat q^2)$, $\hat m_c=m_c/m_b$, $\hat q^2=q^2/m_b^2$), is the function describing charm contribution to 
vacuum polarization. At leading order in $\alpha_s$, one has 
\begin{eqnarray}
g_{cc}(\hat m_c^2,0)=-\frac{8}{9}\log\left(\frac{m_c}{m_b}\right)-\frac{4}{9}\sim 1. 
\end{eqnarray} 
The factorizable charm contribution to the $b\to s l^+l^-$ effective Hamiltonian has the same Lorentz structure as the $C_{9V}$ term 
in the top contribution, Eq.~(\ref{topll}). So, the relative strength of the factorizable charm/top contributions is essentially 
determined by the ratio of the 
Wilson coefficients $C_{2}+3C_1$ and $C_{9V}$ (obvioulsly, the top contribution contains also other Lorentz structures, therefore the total 
top contribution is further enhanced; also the charm contribution contains also nonfactorizable pieces to be discussed later; these are proportional to $C_2$).
The combination $C_{2}+3C_1$ depends strongly on the precise value of the low-energy scale 
$\mu\sim m_b$. For instance, at $\mu=m_b$, one finds $C_2+3C_1=-0.3$, while $C_{9V}=-4.21$. Indeed, there is some numerical 
(although not parametric) suppression of the factorizable charm contribution but this suppression is the subject to the precise 
choice of the scale $\mu$. This indicates the importance of higher-order QCD corrections.

As the final step, the calculation of the amplitude of an exclusive FCNC $B$-decay requires the $B$-meson weak decay form factors. 
For instance, for $B\to \gamma l^+l^-$ decay, one needs the $B\to\gamma$ transition form factors of the $b\to s$ quark currents 
\begin{eqnarray}
\label{ffs}
\nonumber
\langle
\gamma^* (k,\,\epsilon)|\bar s \gamma_\mu\gamma_5 b|\bar B_s(p)\rangle 
&=& 
i\, e\,\epsilon^*_{\alpha}\, \left ( g_{\mu\alpha} \, k'k-k'_\alpha k_\mu \right )\,\frac{F_A(k'^2,k^2)}{M_{B_s}}, 
\\
\langle \gamma^*(k,\,\epsilon)|\bar s\gamma_\mu b|\bar B_s(p)\rangle
&=& 
e\,\epsilon^*_{\alpha}\,\epsilon_{\mu\alpha k' k}\frac{F_V(k'^2,k^2)}{M_{B_s}},   
\\
\langle\gamma^*(k,\,\epsilon)|\bar s \sigma_{\mu\nu}\gamma_5 b|\bar B_s(p) 
\rangle\, k'^{\nu}
&=& 
e\,\epsilon^*_{\alpha}\,\left( g_{\mu\alpha}\,k'k- k'_{\alpha}k_{\mu}\right)\, 
F_{TA}(k'^2, k^2), 
\nonumber
\\
\langle
\gamma^*(k,\,\epsilon)|\bar s \sigma_{\mu\nu} b|\bar B_s(p)\rangle\, k'^{\nu}
&=& 
i\, e\,\epsilon^*_{\alpha}\epsilon_{\mu\alpha k' k}F_{TV}(k'^2, k^2).
\nonumber 
\end{eqnarray}
Let us emphasize that the form factors above determine the $B\to \gamma l^+l^-$ amplitudes of both $H_{\rm eff}^{b\to sl^+l^-}({\rm top})$ 
of Eq.~(\ref{topll}) and $H_{\rm eff}^{b\to sl^+l^-}({\rm charm,fact})$ of Eq.~(\ref{charmll}). 
The peculiar feature of the factorizable charm contribution 
is its proportionality to the combination $(3C_1+C_2)$ which is strongly sensitive to the precise value of the scale $\mu$. 
This property indicates the importance of higher-order QCD corrections. 
The latter should be synchronously taken into account 
in the Wilson coefficients and in the $B$-decay form factors; this is however a very subtle and difficult problem.

\subsubsection{Nonfactorizable charm contribution}
In addition to factorizable charm contributions, one needs to take into account nonfactorizable charm effects induced by soft gluons 
emitted from the charm-quark loop and absorbed in the $B$-meson loop. The latter are not reduced to the product of charm contribution to the vacuum 
polarization and the meson transition form factors, but have a more complicated structure. 
Nonfactorizable charm contributions to the effective Hamiltonian $H_{\rm eff}^{b\to s l^+l^-}$ should be obtained as the convolution of the octet-octet 
part of the effective $b\to s\bar c c$ effective Hamiltonian
\begin{eqnarray}
\label{charml2}
H_{\rm eff}^{b\to s \bar cc}({\rm octet-octet}) =\, 
\frac{G_{F}}{\sqrt2}\,V_{cb}V^*_{cs}\, \frac{C_2}{6}
\left(\bar s \gamma_{\mu}\left (1\, -\,\gamma_5 \right)T^L b\right) \,
\left(\bar c \gamma_{\mu}\left (1\, -\,\gamma_5 \right)T^L c\right),  
\end{eqnarray} 
with the electromagnetic 
\begin{eqnarray}
ie Q_c(\bar c\gamma_\alpha c) A_\alpha^{e.m.}
\end{eqnarray} 
and the strong 
\begin{eqnarray}
ig_s(\bar c\gamma_\beta T^N c) A^N_\beta
\end{eqnarray} 
vertices of the charm-quark field. 
Here $A_\alpha^{e.m.}$ and $A^N_\beta$ are the photon and the gluon field, 
respectively; $T^L=\lambda^L/2$, with $\lambda^L$ the Gell-Mann matrices, the indices $L,N=1,\dots,8$. 

Nonfactorizable charm contributions are governed by a large Wilson 
coefficient $C_2$. So, nonfactorizabe and factorizable charm-loop effects are expected to be of the same order of magnitude. 
Comparing the patterns of broad charmonia measured in $l^+l^-$ collisions 
and in $B\to (K,K^*) l^+l^-$ decays, one concludes that in the charmonia region, nonfactorizable gluon exchanges are indeed at least 
equivalently important as factorizable charm effects.

A number of theoretical analyses of nonfactorizable effects induced by charm-quark contributions has been published in 
the literature: an effective gluon-photon local operator describing the charm-quark loop has been 
calculated in \cite{voloshin} for the real photon as an expansion in inverse charm-quark mass $m_c$ and applied to 
inclusive $B\to X_s\gamma$ decays; Ref.~\cite{buchalla} obtained a nonlocal effective gluon-photon operator for the 
virtual photon and applied it to inclusive $B\to X_s l^+l^-$ decays. 
In \cite{khod1997} nonfactorizable corrections in exclusive FCNC $B\to K^*\gamma$ decays using local OPE have been studied; 
in \cite{zwicky1,zwicky3}, these corrections have been analyzed with light-cone sum rules using local OPE for the photon-gluon 
operator and three-particle light-cone distribution amplitudes of $K^*$-meson.

As emphasized in \cite{voloshin,buchalla,ligeti,paz,hidr}, local OPE for the charm-quark loop leads to a power 
series in $\Lambda_{\rm QCD} m_b/m_c^2$ (see Appendix A). This parameter is of order unity for the physical masses of $c$- and $b$-quarks and 
thus corrections of this type require resummation. The authors of \cite{hidr} derived a different form of the nonlocal 
photon-gluon operator compared to \cite{buchalla} and evaluated its effect at small values of $q^2$ 
($q$ momentum of the lepton pair) making use of light-cone 3-particle DA (3DA) of the $B$-meson with the aligned arguments,
$\langle 0|\bar s(y)G_{\mu\nu}(uy) b(0)|B_s(p)\rangle$. In \cite{mk} it was shown that the consistent resummation 
of $\left(\Lambda_{\rm QCD} m_b/m_c^2\right)^n$ terms requires the generic 3DA of the $B$-meson, 
$\langle 0| \bar s(y)G_{\mu\nu}(x) b(0)|B_s(p)\rangle$, with non-aligned coordinates. 

In the next sections we discuss in detail the contributions of charm to the amplitudes of 
excusive FCNC $B$-decays. To avoid technical complication related to a Lorentz/spinorial structure of the amplitudes, we illustrate
the calcuation using a field theory with scalar quarks/gluons; the generalization to 
QCD is straightforward. We nevertheless keep the QCD notations for quark fields and assume the following hierarchy of quark masses 
$m_b\gg  m_c \gg m_s\sim \Lambda_{\rm QCD}$, and adopt the counting scheme in which the parameter $\Lambda_{\rm QCD}m_b/m_c^2$ is kept of order unity.

\section{Factorizable charm-loop effects}
\begin{figure}[h]
\centerline{\includegraphics[width=5cm]{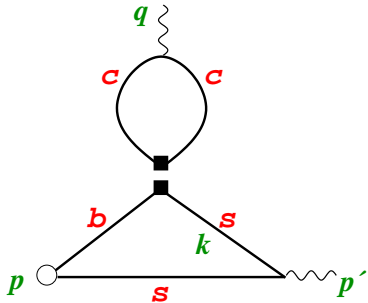}}
\caption{\label{Fig:1}
Factorizable contribution of the charm loop to the amplitude of an FCNC $B$-decay. 
The upper part of the diagram is the charm contribution to the vacuum polariization. 
The lower part of the diagram describes the $B$-meson decay induced by a FCNC $b\to s$ weak current. 
It is determined by the same form factors that describe the amplitude of an FCNC $B$-decay induced ba the top-quark 
contribution in the loop. The solid squares denote the color singlet-singlet four-quark operaotrs.} 
\end{figure}

Factorizable charm-loop contributions are reduced to the product of the $B$-meson weak transition form factors 
induced by bilinear $b\to s$ weak quark currents and the charm contribution 
to vacuum polarization, Fig.~\ref{Fig:1}. 
The latter quantity may be calculated in perturbation theory far below the charm quark threshold, $q^2\ll 4m_c^2$. 
Above the charm threshold, perturbative QCD calculations are not applicable, and approaches based on hadron degrees of freedom 
should be used. However, vacuum polarization can be measured experimentally in a broad range of $q^2$, including the region 
above the charm threshold.

The form factor describing a FCNC transition $B_s\to \gamma^*(q)\gamma^*(p')$ is defined as follows:
\begin{eqnarray}
\label{ff1}
F(q^2,p'^2)&=&
i\int dx e^{iqx}\langle 0|T(\bar s(x)s(x),\bar s(0)b(0))|B_s(p)\rangle.
\end{eqnarray}
Here $p=q+p'$ and $p^2=M_B^2$, and in this formula the quark fields are Heisenberg field operators in QCD. 
By expanding these operators in powers of $\alpha_s$, one generates the corresponding expansion of the form factor. 

\subsection{Leading-order contribution}
\begin{figure}[b]
\centerline{\includegraphics[width=6cm]{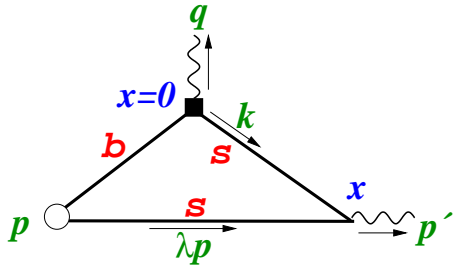}}
\caption{\label{Fig:2}
A schematic calculation of the $B$-decay form factor in QCD: 
the leading QCD contribution given in terms of the $B$-meson quark-antiquark vertex function 
$\langle 0 | \bar s(x)b(0) | B_s(p)\rangle$. The $s$-quark line between the points $x=0$ and $y$ denotes the $s$-quark propagator, whereas 
the $b$-quark line and the $s$-quark line originating from the $B$-meson vertex, 
do not correspond to the propagators of these particles. The latter denotes the operators $b(0)$ and $\bar s(y)$, respectively.
The light-cone configuration of the 2DA [i.e., $y^2=0$] provides the leading contribution to the form factor; 
deviations from the light cone lead to the contributions suppressed as $\Lambda_{\rm QCD}/m_b$.} 
\end{figure}

At leading order in $\alpha_s$, the form factor is given by the diagram of Figure \ref{Fig:2}. 
This contribution to the form factor may be represented as follows:
\begin{eqnarray}
\label{ff10}
F(q^2,p'^2)&=&
i\int dx e^{iqx}\langle 0|T(\bar s(x)b(x),\bar s(0)s(0))|B_s(p)\rangle\nonumber\\
&=&\frac{1}{(2\pi)^4}\int dx e^{iqx}dk e^{-ikx}\frac{\langle 0|\bar s(x)b(0)|B_s(p)\rangle}{m_s^2-k^2-i0}.
\end{eqnarray}
Here the $B$-meson Bethe-Salpeter wave function, $\langle 0|\bar s(x)b(0)|B_s(p)\rangle$, 
depends on two variables $x^2$ and $xp$. One can parametrize it by a Fourier transform in variable $xp$ and perform 
the Taylor expansion in variable $x^2$:   
\begin{eqnarray}
\label{2DA}
\langle 0|\bar s(x)b(0)|B_s(p)\rangle
=\int\limits_0^1 d\xi e^{-ipx \xi}\left\{ \phi_0(\xi)+x^2 \phi_1(\xi)+\dots\right\}
\end{eqnarray}
Because of the general properties of Feynman diagrams, the integration in the variable $\xi$ runs from $0$ to $1$. 
We now insert this expansion in Eq.~(\ref{ff1}) and study the relative size of the different contributions generated by the expansion (\ref{2DA}). 

\noindent $\bullet$ The term proportinal to $\phi_0$ in (\ref{2DA}) corresponds to $x^2=0$ and thus describes the light-cone configuration  
of the quarks inside the $B_s$-meson. Its contribution to the form factor (\ref{ff1}) is easy to calculate:  
\begin{eqnarray}
\label{fflc}
F(q^2,p'^2)=
\frac{1}{(2\pi)^4}\int \frac{dx e^{iqx}\phi_0(\xi)d\xi e^{-i \xi px} e^{-ikx}dk}{m_s^2-k^2-i0}=
\int\limits_0^1 \frac{d\xi \,\phi_0(\xi)}{m_s^2-(q-\xi p)^2}
\end{eqnarray}
Taking into account that $(p-q)^2=p'^2$, and thus $2qp=p^2-q^2-p'^2$, we obtain 
\begin{eqnarray}
k^2=q^2(1-\xi)-\xi(1-\xi)M_B^2+p'^2\xi.
\end{eqnarray}
We now have to take into account the crucial property of the light-cone two-particle distribution amplitude (2DA) $\phi_0(\xi)$ 
of the heavy meson 
$B_s$: since the $b$-quark is heavy, it carries the major part of the $B$-meson momentum, such that $\phi_0(\xi)$ is 
peaked near $\xi\sim \Lambda_{\rm QCD}/m_b$. We then find that $k^2\sim -\Lambda_{\rm QCD}m_b$, i.e.,  
the propagating $s$-quark is highly virtual, and therefore the perturbative expression for its propagator is well justified.
\begin{eqnarray}
F(q^2,p'^2)=
\int\limits_0^1 \frac{\phi_0(\xi)d\xi}{m_s^2-\left(q^2(1-\xi)-\xi(1-\xi)M_B^2+p'^2\xi\right)}
\end{eqnarray}
\noindent
$\bullet$ We now turn to the $x^2$ terms in the expansion (\ref{2DA}): these terms describe the deviations from the light-cone configuration. 
To calculate these contributions, it is convenient to substitute in Eq.~[\ref{fflc}) $x_\alpha=-i\frac{\partial}{\partial k_\alpha} e^{ikx}$. 
By performing the parts integration, the $k_\alpha$-derivative acts on the $s$-quark propagator. Taking into account that
$k^2\sim -\Lambda_{\rm QCD}m_b$, we find that 
the contribution of a term $(x^2)^n\to (\Lambda_{\rm QCD}/m_b)^n$ compared to LC term. So the integral for the form factor (\ref{ff1}) 
is indeed dominated by the light-cone quark configuration, whereas the deviations from the LC in the BS wave function of the $B$-meson 
are suppressed by powers of a small parameter $\Lambda_{\rm QCD}/m_b$.

\subsection{Next-to-leading-order contribution}
We now calculate $\alpha_s$-corrections generated by the expansion of the $T$-product in Eq.~(\ref{ff1}).
\begin{figure}[b]
\includegraphics[width=6cm]{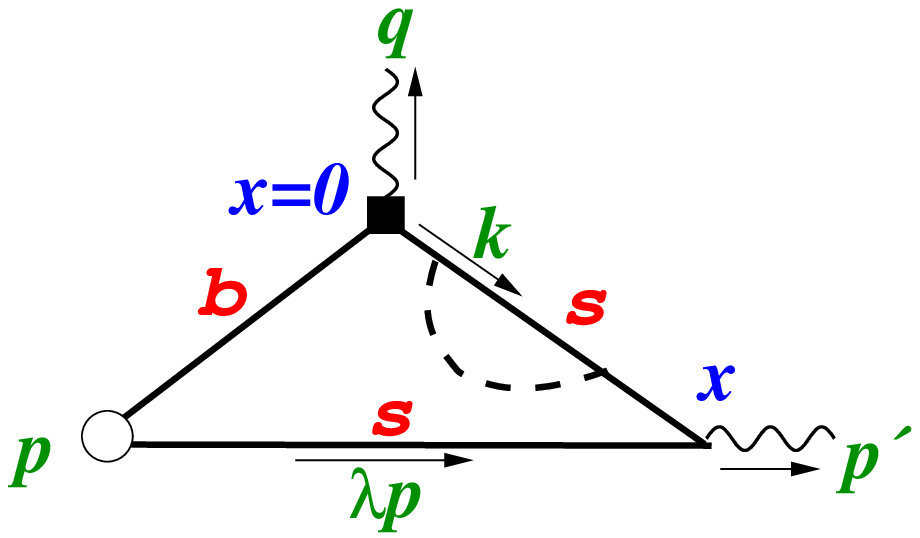} \qquad \includegraphics[width=6cm]{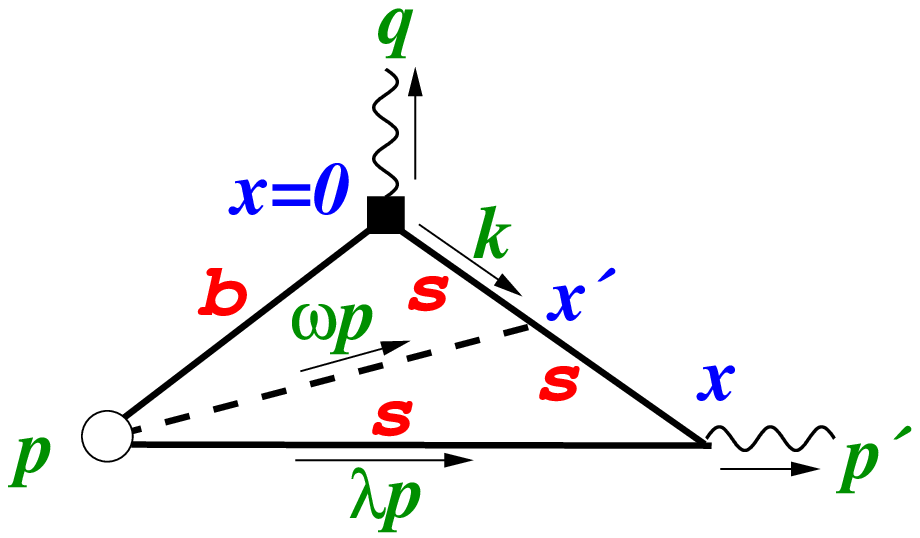}
\\
${}$ \hspace{2cm}(a) \hspace{7cm}  (b)
\caption{\label{Fig:1rad}
The NLO corrections to the $B_s$ weak decay form factor.  
(a) the NLO correction given in terms of the $B$-meson quark-antoquark vertex 
$\langle 0 | \bar s(x) b(0) | 0\rangle$. 
(b) soft-gluon correction to the leading-order contribution given in terms of the 3-particle DA 
$\langle 0 | \bar s (x))G(x')b(0)| 0\rangle$. 
In both cases (a) and (b), the light-cone configurations of the 2DA and the 3DA 
[i.e., $y^2=0$ in the case (a) and $x^2=x'^2=(x-x')^2=0$ in the case (b)] 
provide the leading contribution to the form factor; deviations from the light cone in both cases 
lead to the contributions suppressed as $\Lambda_{\rm QCD}/m_b$. Recall that the quark and the gluon lines attached 
to the $B_s$ meson vertex do not represent the propagators of the corresponding particls, but denote the field operators 
$\bar s(x)$, $G(x')$, and $b(0)$.} 
\end{figure}
The diagram of Fig.~\ref{Fig:1rad}(a) corresponds to the radiative correction to the $s$-quark 
propagator. As we have shown above, the $s$-quark is highly virtual, so the calculation of the radiative 
correction is straightforward. We do not give any detail 
of this calculation but just notice that the corresponding contribution is again given in terms of the two-particle BS amplitude 
of Eq.~(\ref{2DA}). Similar to the LO contribution discussed above, the LC term in the expansion of Eq.~(\ref{2DA}) dominates the 
$B_s$-decay amplitude, whereas the terms containing powers of $(x^2)^n$ lead to power-suppressed 
terms $\sim (\Lambda_{\rm QCD}/m_b)^n$ in the $B_s$-decay amplitude. 

Another correction to the $B_s$ decay amplitude shown in Fig.~\ref{Fig:1rad}(b) has a different structure and needs a more complicated three-particle 
quark-antiquark-gluon distribution amplitude (3DA) $\langle 0|\bar s(y)G(x')b(0)|B_s(p)\rangle$ for its calculation. A detailed discussion 
of 3Da will be given in the next Section. Here we just emphasize, that the LC term in 3DA leads to the dominant part of the $B_s$ decay 
amplitude, whereas all the terms in 3DA containing powers of $(x'^2)^n$, $(y^2)^n$, and $((x'-y)^2)^n$,  lead to the contributions to the $B_s$ 
decay amplitude, that are suppressed by powers of $(\Lambda_{\rm QCD}/m_b)^n$. So the situation is very similar to the case of the 
diagram of Fig.~(\ref{Fig:1}. This is due to the fact that the factorizable contributions discussed in this Section contain only one fermion loop 
containing the heavy quark $b$. We shall see in the next Section that the situation is different in the case of nonfactorizabe contributions.
In the latter case, the amplitude involves two quark loops with different masses, $m_b$ and $m_c$, and this fact changes the suppression of 
the contributions to the $B_s$ amplitude, generated by the off-LC terms in the 3DA of $B_s$.

\section{Nonfactorizable charm-loop effects}
\label{Sec_model}
The charm contribution to the amplitude of an FCNC $B$-decay reads 
\begin{eqnarray}
\label{Apqoriginal}
A(p,q)=i\,\int dz e^{i q z}\langle 0|T\{\bar c(z)c(z),\,\bar s(0)s(0)\}|B_s(p\rangle, 
\end{eqnarray}
where quark fields are understood as Heisenberg field operators in a theory that involves weak and strong interactions. 
Our goal is to study nonfactorizable corrections due to a soft-gluon exchange 
between the charm-quark loop and the $B$-meson loop. To the lowest order, the corresponding amplitude is given by 
the diagram of Fig.~\ref{Fig:1}:  
\begin{eqnarray}
\label{Apqscalar}
&&A(p,q)=i\,\int dz e^{iq z}\nonumber\\
&&\qquad \times\langle 0|T\{\bar c(z)c(z),
\,i\int dy' \,L_{\rm weak}(y'),
\,i\int dx\, L_{Gcc}(x), 
\,\bar s(0)s(0)\}|B_s(p\rangle. 
\end{eqnarray}
\begin{figure}[b]
\centerline{\includegraphics[width=6cm]{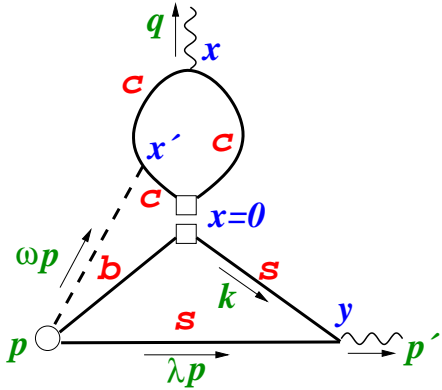}}
\caption{\label{Fig:3}
One of the diagrams describing the nonfactorizable gluon exchange. 
Dashed line corresponds to gluon; $q$ and $\kappa=-\omega p$ are the 
momenta outgoing from the charm-quark loop; the momentum $q'=q+\kappa=q-\omega p$ is emitted from the $b\to s$ vertex. 
Another diagram, equal to the one shown in the figure, corresponds to the gluon attached to the right $c$-quark line 
in the upper loop. The empty squares denote the color octet-octet four-quark operators.} 
\end{figure}
The effective Lagrangian that mimics weak four-quark interaction is chosen in a simple form  
\begin{eqnarray}
L_{\rm weak}=\frac{G_F}{\sqrt2}\, \bar s b\,\bar c c,
\end{eqnarray}
while the interaction of the scalar gluon field $G(x)$ and the scalar $c$-quarks is taken to be   
\begin{eqnarray}
L_{\rm Gcc}=G(x)\,\bar c(x) c(x). 
\end{eqnarray}
We can use the gluon field in momentum representation, which is related to the gluon field in coordinate representation as follows
\begin{eqnarray}
G(x)=\frac{1}{(2\pi)^4}\int d\kappa \,\tilde G(\kappa)\,e^{i\kappa x},\quad \tilde G(\kappa)=\int dx \,G(x)\,e^{-i\kappa x}. 
\end{eqnarray}
Then the effective operator describing the gluon emission from the charm quark loop may be written as  
\begin{eqnarray}
\label{t1}
{\cal O}(q)=\int d\kappa\, \tilde G(\kappa)\,\Gamma_{cc}(\kappa,q),  
\end{eqnarray}
where $\Gamma_{cc}(\kappa,q)$ stands for the contribution of two triangle diagrams with the charm quark running in the loop. 
The momenta $\kappa$ and $q$ are outgoing from the charm-quark loop, whereas the momentum $q'=q+\kappa$ 
is emitted from the $b\to s$ vertex. $p'$ is the momentum of the outgoing $\bar s s$ current and $p$ is the momentum of the $B$-meson, 
$p=p'+q$. 

In terms of the gluon field operator in coordinate space, we can rewrite (\ref{t1}) as 
\begin{eqnarray}
\label{t2}
{\cal O}(q)=\int d\kappa\, e^{-i\kappa x}dx\,G(x)\Gamma_{cc}(\kappa, q), 
\end{eqnarray}

\subsection{Three-particle antiquark-quark-gluon distribution amplitude of $B$-meson}
By virtue of (\ref{t2}), the amplitude Eq.~(\ref{Apqscalar}) takes the form
\begin{eqnarray}
\label{A}
A(q,p)&=&\frac{1}{(2\pi)^8}\int \frac{dk}{m_s^2-k^2}\int dy e^{-i(k-p')y}\int dx' e^{-i\kappa x'}\nonumber\\
&&\qquad\qquad\times\int d\kappa\, 
\Gamma_{cc}(\kappa, q)\,\langle 0|\bar s(y)G(x') b(0)|B_s(p)\rangle. 
\end{eqnarray}
Here, we encounter the $B$-meson three-particle amplitude with three (non-aligned) arguments, 
for which one may write down the following decomposition: 
\begin{eqnarray}
\label{3DAnew}
\langle 0|\bar s(y)G(x') b(0)|B_s(p)\rangle&=&
\int d\lambda e^{-i \lambda y p}
\int d\omega e^{-i \omega x p}\,\nonumber\\
&&\times\left[\Phi(\lambda,\omega)+O\left(x'^2,y^2,(x'-y)^2\right)\right].   
\end{eqnarray}
Here $\lambda$ and $\omega$ are dimensionless variables. Making use of the properties of Feynman diagrams, 
one may show that they should run from 0 to 1. However, if one of the meson constituents is heavy, it carries the 
major fraction of the meson momentum and as the result the function $\Phi(\lambda,\omega)$ is strongly peaked 
in the region  
\begin{eqnarray}
\label{peaking}
\lambda, \omega=O(\Lambda_{\rm QCD}/m_b).  
\end{eqnarray}
So, effectively one can run the $\omega$ and $\lambda$ integrals from 0 to $\infty$; such integration limits in fact 
emerge in the DAs within heavy-quark effective theory \cite{hidr,japan}. We emphasize that 
for the results presented below only the peaking of the DAs in the region (\ref{peaking}) is essential. 
Notice also that the function $\Phi(\lambda,\omega)$ in (\ref{3DAnew}) coincides with the same 
function that appears in the ``standard'' 
3-particle distribution amplitude with the aligned arguments, $x=u y$, discussed in \cite{japan}.

\subsection{Light-cone contribution}
First, let us calculate the contribution to $A(q,p)$ from the term that corresponds to $x'^2=y^2=(x'-y)^2=0$ 
in the 3DA (\ref{3DAnew}). This is very easily calculable: by inserting (\ref{3DAnew}) into (\ref{A}) 
we can perform the $x-$ and $y-$integrals
\begin{eqnarray}
\label{xyint}
&&\int dx'\to\delta(\kappa +\omega p), \qquad \nonumber\\
&&\int dy\to\delta(k+\lambda p-p'). 
\end{eqnarray}
Next, the $\delta$-functions above kill the integrals over $k$ and $\kappa$, and we find
\begin{eqnarray}
\label{Aqp}
A(q,p)=\int_0^\infty d\lambda \int_0^\infty d\omega\, \Phi(\lambda,\omega)
\Gamma_{cc}\left(-\omega p, q \right)\frac{1}{m_s^2-(\lambda p-p')^2}.  
\end{eqnarray}
\subsubsection{$s$-quark produced in a weak $b\to s$ transition is highly virtual}
By virtue of the $y$-integration in Eq.~(\ref{xyint}), the $s$-quark propagator has taken the form  
\begin{eqnarray}
m_s^2-(\lambda p-p')^2=m_s^2-\lambda q^2+(1-\lambda)(\lambda M_B^2-{p'}^2).
\end{eqnarray} 
Because of the property of the $B_s$-meson 3DA, the $\lambda$-integral is dominated by the region 
\begin{eqnarray}
\lambda\sim \Lambda_{\rm QCD}/m_b. 
\end{eqnarray}
Therefore, in the bulk of the $\lambda$-integration the virtuality of the $s$-quark is large, $k^2\sim -\Lambda_{\rm QCD}M_B$,  
and the use of the Feynman expression for the $s$-quark propagatorin in Eq.~(\ref{Aqp}) is well justified. 

\subsubsection{Quarks inside the charm-quark loop with soft gluon emission are perturbative}
The charm contribution is described by the three-point function $\Gamma_{cc}(\kappa,q)$, which depends on three invariants: 
$q^2$, $q'^2$ and $\kappa^2$. For our analysis it is important, that in the region 
$q^2\ll 4m_c^2$, $q'^2$ and $\kappa^2$ are also far below the charm threshold, i.e., quarks in the charm loop are perturbative. 

The momentum transferred in the weak-vertex is $q'=q+\kappa=q-\omega p$, such that 
\begin{eqnarray}
q'^2=(q-\omega p)^2=q^2-\omega (1-\omega )M_B^2-q^2\omega +p'^2 \omega =q^2-\omega(1-\omega)M_B^2.
\end{eqnarray}
We now take into account that the $\omega$-integral is dominated by 
\begin{eqnarray}
\omega\sim \Lambda_{\rm QCD}/m_b
\end{eqnarray}
In the region $\omega\sim \Lambda_{\rm QCD}/m_b$ and $q^2\ll 4m_c^2$, one finds $q'^2\sim -\Lambda_{\rm QCD}m_b$. 
Also $\kappa^2=\omega^2 p^2\simeq \Lambda_{\rm QCD}^2$. 
So, all three invariants desribing the triangle charm-quark loop are far below the charm threshold.
One can then use the perturbative Feynman $c$-quark propagators for the calculation of $\Gamma_{cc}(\kappa,q)$, and 
obtains for the sum of two triangle diagrams (with the charm quark running in the loop in two opposite directions) 
the following expression: 
\begin{eqnarray}
\label{tFeyn}
\Gamma_{cc}(\kappa, q)=\frac{1}{8\pi^2}
\int\limits_{0}^{1}du \int\limits_{0}^{1-u}dv \,
\frac{1}{m_c^2-2uv \kappa q -u(1-u)\kappa^2-v(1-v)q^2}. 
\end{eqnarray}

\subsection{Deviations from the light-cone}
We now turn to the calculation of the contributions to $A(q,p)$ generated by terms $\sim x^2,y^2,(x-y)^2$ in the 3DA (\ref{3DAnew}). 
The terms containing powers of 4-vectors $y$ and $x$ in the integral (\ref{A}) can be calculated by parts 
integration. Taking into account the results (\ref{xyint}), we find the following relative contributions of the terms 
containing different powers of the coordinate variables: 
\begin{eqnarray}
\Lambda_{\rm QCD}^2 y^2 &\to& \frac{k^2}{m_b^2}\sim \frac{\Lambda_{\rm QCD}}{m_b},\nonumber\\
\Lambda_{\rm QCD}^2 x'^2&\to& \Lambda_{\rm QCD}^2\frac{q\kappa}{m_c^4}\sim \frac{\Lambda^3_{\rm QCD} m_b}{m_c^4},\nonumber\\
\Lambda_{\rm QCD}^2 x'y&\to& \Lambda_{\rm QCD}\frac{(p'-\lambda p )(q-\omega p)}{m_b m_c^2}\sim \frac{\Lambda_{\rm QCD} m_b}{m_c^2}.
\end{eqnarray}
Clearly, all terms containing powers of $x'^2$ and/or $y^2$ in the 3DA lead to the suppressed contributions to $A(q,p)$ 
and may be neglected within the considered accuracy.  
However, the terms containing powers of $x'y$ lead to the contributions containing powers of ${\Lambda_{\rm QCD}m_b}/{m_c^2}$, 
i.~e., of order unity within the adopted counting rules. The kinematics of the process is thus rather simple: 
the vectors $x'$ and $y$ are directed along the light-cone 
[e.g., $x'$ along the $(+)$ axis, and $y$ along the $(-)$ axis], but the 4-vector 
$(x'-y)$ is obviously not directed along the light cone. Therefore, the full dependence of 3DA on the variable $(x'-y)^2$ 
is needed in order to properly resum all corrections of order $\left(\Lambda_{\rm QCD}m_b/m_c^2\right)^n$.


\section{Conclusions and outlook}
\begin{figure}[h]
\centerline{\includegraphics[width=9cm]{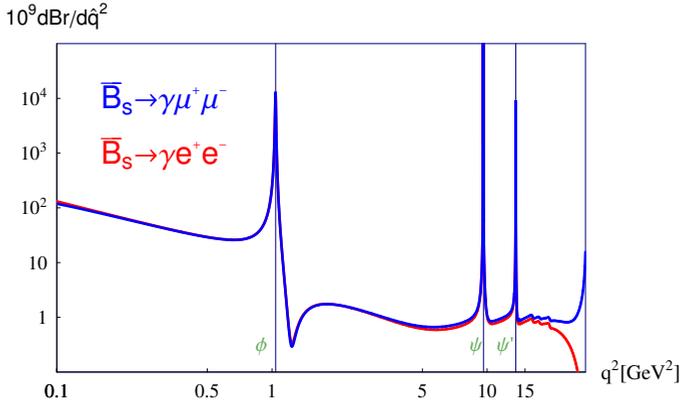}} 
\caption{\label{Fig:BR}
The predicted differential branching ratios $d{\rm Br}/d\hat q^2$ in $B\to \gamma l^+l^-$ from \cite{mnk}. $\hat q^2=q^2/M_B^2$.} 
\end{figure}

We reviewed the general framework for the theoretical description of semileptonic and radiative leptonic 
$B$-decays induced by flavour-changing neutral currents with an emphasis on gaining control over 
QCD effects induced by charming loops, including both factorizable and nonfactorizable contributions. 
The charm-loop effects are known to be small compared to the effects of the $t$-quark in the loop at small values of the dilepton pair 
(although there is no parametric suppression of charm compared to that of top), but dominate the amplitudes of FCNC rare $B$-decays 
in the region of charmonium resonances. As an illustration of the typical magnitude of the contribution of charm, Fig.~\ref{Fig:BR} 
\begin{figure}[t]
\centerline{\includegraphics[width=9cm]{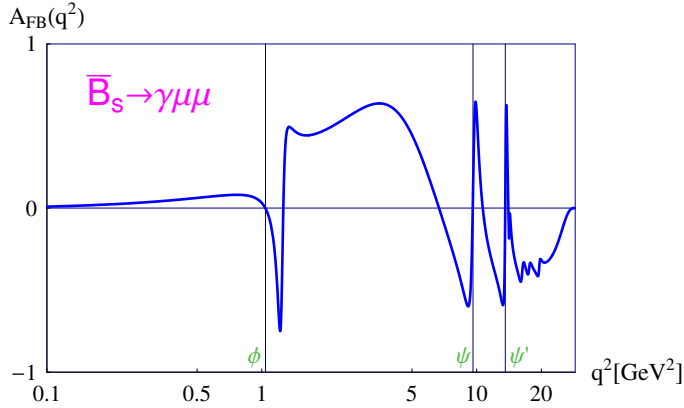}}
\caption{\label{Fig:AFB}
The predicted forward-backward asymmetry $A_{FB}(q^2)$ in $B\to \gamma l^+l^-$ decays from \cite{mnk}.} 
\end{figure}
shows 
the predicted differential 
distribution in $B\to \gamma l^+l^-$ decays from \cite{mnk} where charm is clearly seen to provide the main contribution to the 
branching ratios in the charmonia region. Charm contributions in the charmonia resonance region also strongly influence various observables; 
Fig.~\ref{Fig:AFB} shows the prediction for the lepton forward-backward asymmetry in $B_s\to \gamma l^+l^-$ from \cite{mnk}. 
The contribution of charm away from the charmonia region is expected at the level of a few percent thus 
providing an unpleasant ``noise'' for the search for the potential new physics effects. 
This calls for a better understanding of the charm contributions in FCNC $B$-decays and gaining a 
theoretical control over these contributions.

To put our emphasis on the conceptual aspects, we discussed the case of all scalar particles, 
avoiding in this way conceptually unimportant technical details.
 
Our main conclusions may be formulated as follows: 

\begin{itemize}
\item
The relevant object that arises in the calculation of the nonfactorizable corrections is the three-particle DA (or the vertex function) 
with independent (non-aligned) coordinates: 
\begin{eqnarray}
\label{res1}
\hspace{-.2cm}
\langle 0|\bar s(y)G(x)b(0)|B_s\rangle=
\int d\lambda e^{-i \lambda y p}
\int d\omega e^{-i \omega x p}\left[\Phi(\omega,\lambda)+O\left(x^2,y^2,(x-y)^2\right)\right]. 
\end{eqnarray} 
This object may be viewed as a four-point function and thus depends on five independent variables (the external momentum 
$p$ is on the mass shell, $p^2=M_B^2$). It is convenient to choose the variables $xp$, $yp$, $x^2$, $y^2$, and $(x-y)^2$ as these five 
independent variables. One may parametrize the function by its Fourier transform in the variables $xp$ and $yp$, and to write the 
Taylor series in the variables $x^2$, $y^2$, and $(x-y)^2$. It should be emphasized that the function $\Phi(\omega,\lambda)$ here is precisely 
the same function that parametrizes the standard 3DA with the aligned arguments, $x=uy$, discussed in \cite{japan}.
  
At small $q^2\le m_c^2$, terms of order $x^2$ and $y^2$ in the 3DA of $B_s$ meson yield small corrections to the nonfactorizable 
amplitude of $B$-decay compared to the LC term in the 3DA: 
for terms $O(x^2)$ the suppression parameter is $\Lambda_{\rm QCD}^2/m_c^2$, and for terms $O(y^2)$ the suppression parameter 
is $\Lambda_{\rm QCD}/m_b$. 
However, terms $\sim (xy)^n$ in the 3DA yield the contributions of order $\left(\Lambda_{\rm QCD}m_b/m_c^2\right)^n$ 
in the $B_s$ decay amplitude. These contributions are of $O(1)$ in our counting scheme and should be properly resummed. (Recall 
that the difference between the local OPE \cite{voloshin} and the light-cone OPE \cite{hidr} is of the same order). 
We emphasize that the knowledge of merely the light-cone 3DA corresponding to $x^2=0$, $y^2=0$ and $(x-y)^2=0$,
is not sufficient: While the LC 3DA allows one to resum a part of the large corrections of the order $\Lambda_{\rm QCD} m_b/m_c^2$, 
some terms of the 
same order remain unaccounted for. A consistent resummation of these corrections 
requires the knowledge of the full dependence of $\langle 0|\bar s(y)G(x)b(0)|B_s(p)\rangle$ on the variable $(x-y)^2$.
\footnote{As emphasized to us by Danny van Dyk, the actual suppression parameter of those contributions to the amplitude of FCNC $B_s$ decay 
induced by the $(x-y)^2$ terms in 3DA vs the LC terms in 3DA is $\Lambda_{\rm QCD}m_b/(4m_c^2-q^2)$. 
For $|q^2|\sim m_c^2$, this agrees with our claim. If one considers the region of $q^2\sim -m_b^2$, then the  
off-LC effects turn out to be suppressed. However, whether or not one can avoid the calculation of the off-LC effects 
by performing an extrapolation from the region $q^2\sim -m_b^2$ to the region $q^2\sim m_c^2$, requires further detailed analysis.} 

The kinematics of the process looks simple: the 4-vectors $x$ and $y$ are directed along the light-cone 
[e.g., $x$ along the $(+)$-axis, and $y$ along the $(-)$-axis], but the 4-vector 
$x-y$ is obviously not directed along the LC; therefore, the full dependence of the 3DA (\ref{res1}) 
on the variable $(x-y)^2$ is needed in order to properly resum corrections of order $\left({\Lambda_{\rm QCD}m_b}/{m_c^2}\right)^n$. 
\item
We point out that {\it nonfactorizable} soft-gluon corrections to the amplitudes of FCNC $B$-decays have 
qualitatively different features compared to soft-gluon corrections to the $B$-meson form factors
$\langle 0|T \{\bar s s(z),\, \bar s b(0)\}|B(p)\rangle$, Fig.~\ref{Fig:2}. In the latter case, the $B_s$ decay amplitudes are dominated 
by the light-cone configurations of quarks (and gluons) in the $B_s$ meson two- and three-particle vertex functions (DAs). 
The deviations from the LC lead to small $\sim \Lambda_{\rm QCD}/m_b$ corrections to the amplitudes. 
\item
In general, when considering non-factorizable gluon corrections in meson-to-vacuum 
transition amplitudes of the type $\langle 0|T \{j_1(z) j_2(0)\}|B\rangle$ one encounters two distinct kinds of processes: 

(i). The amplitude of the process involves only one quark loop with the valence $b$-quark. An example of this kind is  
the contribution to the $B_s\to \gamma$ form factor due to soft-gluon exchange, Fig.~\ref{Fig:2}. 
In this case, nonfactorizable soft-gluon correction is light-cone dominated, i.e., may be expressed via the LC 3DA 
of the initial $B$-meson \cite{lcsr1,kou,lcsr2}. The contributions to the form factor related to powers of $x^2,y^2,(x-y)^2$ in the $B$-meson 
3DA are suppressed by powers of the small parameter $\Lambda_{\rm QCD}/m_b$. 

(ii). The amplitude of the process involves two separate quark loops (one quark-loop involving valence quarks of 
the initial and the final mesons and another quark loop that emits the external boson). In this case, the soft gluon from the 
initial heavy meson vertex is absorbed by a quark in a different loop. In this case, the description 
of non-factorizable soft-gluon corrections requires the full three-particle DA with non-aligned coordinates of the type 
of (\ref{res1}). Non-factorizable corrections to FCNC decays due to $c$- or $u$-quark loops belong to this kind of processes. 
\end{itemize}

\section*{Acknowledgments}
Many thanks are due to Danny van Dyk for lively and interesting discussions of nonfactorizable effects in FCNC $B$-decays. 
I gratefully acknowledge financial support from RFBR and CNRS under joint CNRS/RFBR Grant PRC Russia/19-52-15022 
and from the Austrian Science Fund (FWF), Project No. P29028. This research was supported in part by the Munich Institute for 
Astro- and Particle Physics (MIAPP) of the DFG Excellence Cluster ORIGINS within the Program ``Deciphering Strong-Interaction 
Phenomenology through Precision Hadron-Spectroscopy''.  

\newpage
\appendix

\section{Local vs non-local OPE}
In this Appendix we discuss the relationship between local and non-local OPE for the amplitude of an FCNC weak decay 
of the $B_s$ meson.

To implement local OPE for the nonfactorizable part of the FCNC $B_s$-decay amplitude, one needs to perform the Taylor expansion 
of the gluon field operator $G(x)$: 
\begin{eqnarray}
G(x)=G(0)+ x_\alpha \partial^\alpha G+\dots. 
\end{eqnarray}
\noindent 
1. The leading contribution comes from $G(0)$ term, and it can be given in terms of the 3DA with the equal coordinates of the gluon 
and the $b$-quark fields: 
\begin{eqnarray}
\label{app1}
\langle 0|\bar s(y)G(0)b(0)|B(p)\rangle|_{x=0}=
\int d\lambda e^{-i \lambda y p}\int d\omega \,\left[\Phi(\lambda,\omega)+O(y^2)\right].
\end{eqnarray}

\noindent 
2.
Let us now consider the contribution of $x_\alpha \partial^\alpha G(0)$. 

\noindent $\bullet$ 
The $x_\alpha$ factor can be generated by $-i\frac{\partial}{\partial q_\alpha} e^{i q x}$ and, after performing the parts integration,  
leads to terms proportional to $\frac{q^\alpha}{m_c^2}$. 

\noindent $\bullet$ 
Thus, after the integration, the term containing $\partial^\alpha G(0)$ takes the form 
\begin{eqnarray}
\label{der}
\frac{\partial}{\partial x_\alpha}\langle 0|\bar s(y)G(x) b(0)|B(p)\rangle|_{x=0}&=&
-ip_\alpha\int d\lambda e^{-i \lambda y p}\int d\omega \,\omega \Phi(\lambda,\omega)+C_2 \Lambda_{\rm QCD}x_\alpha\nonumber \\
&=&C_1 p_\alpha \frac{\Lambda_{\rm QCD}}{m_b}+C_2 \Lambda_{\rm QCD}x_\alpha
\end{eqnarray}
The term $C_2$ arises when the derivative acts on $x^2$ and $xy$ terms in the full off-LC 3DA. 

\noindent 
3. As the next step, we must plug the derivative term (\ref{der}) into the general representation for the nonfactorizable part of 
the $B_s$ amplitude (\ref{A}). Now, let us compare the contributions to the amplitude (\ref{A}), generated by the terms 
$\partial_\alpha G(0)$ over the $G(0)$ in the 3DA of $B_s$ meson. The leading part in the ratio of the $\partial_\alpha G(0)$
 over the $G(0)$ contributions to the amplitude arises 
when $q_\alpha$ contracts with the term $\sim p_\alpha$ and reads  
\begin{eqnarray}
\frac{qp \Lambda_{\rm QCD}}{m_b m_c^2}\sim \frac{M_B\Lambda_{\rm QCD}}{m_c^2}\sim 1.
\end{eqnarray}
For the realistic case of $c-$ and $b$-quarks, and within the adopted counting scheme $\frac{M_B\Lambda_{\rm QCD}}{m_c^2}\sim 1$, 
there is no suppression of the derivative-term contribution. So we conclude that the local OPE does not provide a hierarchy of  
contributions given by different operators according to their dimension. This means that a summation of infinitely many local 
operators is necessary in order to properly account for the terms of the type $(\frac{M_B\Lambda_{\rm QCD}}{m_c^2})^n$. Precisely this is 
is done by considering the non-local OPE.


\end{document}